\documentclass[prl,floatfix,twocolumn]{revtex4}
\usepackage[T1]{fontenc}
\usepackage{graphicx}
\usepackage{hyperref}
\usepackage{times,amssymb,amsmath}

\newcommand{\bfp}{\ensuremath{\boldsymbol{p}}}
\newcommand{\bfq}{\ensuremath{\boldsymbol{q}}}
\newcommand{\bfr}{\ensuremath{\boldsymbol{r}}}
\newcommand{\exhatH}{\langle \hat{H} \rangle}

\begin{document}

\title{Localized electron wave packet description of chemical bond and excitation:
Floating and breathing Gaussian with valence-bond coupling}

\author{Koji Ando\footnote{E-mail: ando@kuchem.kyoto-u.ac.jp}}

\address{Department of Chemistry, Graduate School of Sciences, Kyoto University,
Sakyo-ku, Kyoto 606-8502, Japan}

\begin{abstract}
A model of localized electron wave packets (WPs)
with variable position and width (floating and breathing)
that are
spin-coupled as per the valence-bond theory
is presented.
It produces accurate potential energy curves of LiH
in the ground singlet and triplet states. 
Quantization 
in a mean-field approximation 
of the motion of a WP
that corresponds to the Li 2s electron
generates semi-quantitative potential energy curves of low energy 
excited states.
Real-time semiquantal dynamics 
of the WP
induced by an intense laser pulse 
gives high-harmonic generation spectra that capture qualitative features
of a higher-level wave function calculation.
\end{abstract}

\maketitle

\section{Introduction}

The valence-bond (VB) theory provides many useful concepts for understanding 
formation and reactivity of chemical bond 
\cite{Pauling60,CooperD86,Warshel91,Truhlar06,Shaik08,Wu11,Small11,Karadakov12}.
The standard VB method is
based on the linear combination of
atomic orbitals (LCAO) where the AOs are clumped at the nuclear centers
and the orbital exponents are fixed at predetermined values. 
Variation of the electronic wave function is described by superposing 
a number of AOs 
of different orbital exponents and angular momenta.

The LCAO model is useful particularly for static problems
but is not imperative.
We have been studying in recent years
a model of floating and breathing
(variable position and width)
Gaussian wave packets (WPs) with 
non-orthogonal VB coupling \cite{Ando09,Ando12,Ando16}.
It was designed to be combined with nuclear WPs in
a semiquantal
molecular dynamics (MD) simulation \cite{Kim14,Kim14prb,Kim15,Kim16}.
In order to take account of the anti-symmetry of many-electron wave functions 
retaining the localized WP picture, the VB model seemed appropriate.
The localized WP model was envisaged to have merits for studying 
real-time electron dynamics.

In our previous studies,
the WP-VB model was found to give
reasonably accurate potential energy surfaces of
the electronic ground states of small molecules 
such as H$_2$, LiH, BeH$_2$, H$_2$O, and NH$_3$
with the minimal numbers of electron WPs \cite{Ando09,Ando12,Ando16}.
The accuracy came from the flexibility 
to describe the static correlation by the VB coupling,
the dynamic correlation by the WP breathing, and the polarization
by the WP floating.
The electronic excited states have been also examined by
quantizing the motion of WP center \cite{Ando16}
in an idea similar to the propagator theory \cite{McWeeny92}.
Furthermore, the nuclear and electronic WPs were combined
(with static electrons in the Born-Oppenheimer approximation)
to the semiquantal MD simulation of liquid hydrogen, 
which gave quantitative diffusion constant and viscosity 
at temperatures below 25 K \cite{Kim14,Kim14prb,Kim15,Kim16}.

This work extends the study on LiH in Ref. \cite{Ando16}
to examine the potential
energy surfaces for the electron WP motion
and the real-time dynamics induced by an intense laser pulse.
The study of real-time electron dynamics is an emerging field of recent research
with the advent of attosecond time-resolved laser spectroscopy \cite{Kling08,Krausz09}.

In Sec. 2, the theory is outlined. 
In Sec. 3.1, the potential energy curves for the electron WP motion
are examined.
In Sec. 3.2, the laser-induced electron dynamics 
and the emission of high-harmonic generation spectra
are studied.
Section 4 concludes.

\section{Theory}

As in the standard VB methods,
the electronic wave function is assumed to be an antisymmetrized product 
of spatial
and spin functions,
\begin{equation}
\Psi(1,\cdots,N)={\cal A}[\Phi(\bfr_{1},\cdots,\bfr_{N})\Theta(1,\cdots,N)],
\end{equation}
with the spatial part modeled by a product of one-electron functions,
\begin{equation}
\Phi(\bfr_{1},\cdots,\bfr_{N})=\phi_{1}(\bfr_{1})\cdots\phi_{N}(\bfr_{N}) .
\end{equation}
In contrast with the conventional VB methods that employ
atomic orbitals clumped at the nuclear centers, 
we apply floating and breathing
spherical Gaussian WPs of a form \cite{Arickx86,Pattanayak94},
\begin{equation}
\phi(\bfr,t)
=(2\pi\rho_{t}^{2})^{-\frac{3}{4}}
\exp[-\gamma_{t}|\bfr-\bfq_{t}|^{2}
+ {i}\bfp_{t}\cdot(\bfr-\bfq_{t})/\hbar] ,
\label{eq:wpbasis}
\end{equation}
in which the complex coefficient
\begin{equation}
    \gamma_{t}=\frac{1}{4\rho_{t}^{2}}-\frac{i}{2\hbar}\frac{\pi_{t}}{\rho_{t}}
\end{equation}
contains the time-dependent WP width $\rho_t$ 
and its conjugate momentum $\pi_t$ (see Eq. (\ref{eq:EOMrhopi}) below).
$\bfq_{t}$ and $\bfp_{t}$ denote the position of the WP center and its conjugate momentum.
The spin part $\Theta(1,\cdots,N)$ consists of the spin eigenfunctions. 
In this work, we employ a single configuration of the perfect-pairing form,
\begin{equation}
\Theta =
\theta(1,2)
\cdots\theta(N_{p}-1,N_{p})\alpha(N_{p}+1)\cdots\alpha(N) ,
\end{equation}
with
$\theta(i,j)=(\alpha(i)\beta(j)-\beta(i)\alpha(j))/\sqrt{2}$
for paired $N_p$ electrons.

The ground state wave function and energy in a given spin multiplicity
is computed by optimizing the WP center and width, 
$\bfq$ and $\rho$,
to minimize the energy expectation, 
$E=\langle\Psi|\hat{H}|\Psi\rangle/\langle\Psi|\Psi\rangle$,
with the momentum variables $\bfp$ and $\pi$ nullified \cite{Tsue92}.
For the excited states, the standard method would be the 
configuration-interaction (CI) of multiple VB configurations \cite{McWeeny92}.
Alternatively, we are working on a propagator theory
with the coherent-state path-integral representation \cite{Ando14}, 
exploiting the fact that the Gaussian WP is a representation of the (over)complete
coherent-state \cite{Tsue92,Kuratsuji81,Klauder85}.
In Sec. 3, a simplified calculation based on this idea 
with the single-particle mean-field approximation 
is presented.

In the single particle approximation,
the equations of motion for the WP center and width 
have a simple Hamiltonian form \cite{Arickx86,Pattanayak94,Ando04hext,Prezhdo06},
\begin{equation}
    \dot{q} = \frac{\partial \exhatH}{\partial p},
\hspace*{1em}
    \dot{p} = - \frac{\partial \exhatH}{\partial q},
\label{eq:EOMqp}
\end{equation}
\vspace*{-1em}
\begin{equation}
    \dot{\rho} = \frac{\partial \exhatH}{\partial \pi},
\hspace*{1em}
    \dot{\pi} = - \frac{\partial \exhatH}{\partial \rho} ,
\label{eq:EOMrhopi}
\end{equation}
with the Hamiltonian function
\begin{equation}
\exhatH
= \frac{p^2}{2m} + \frac{\pi^2}{2m} +
  \frac{\hbar^2}{8 m \rho^2} + \langle V \rangle ,
\label{eq:Hext}
\end{equation}
where $m$ is the mass of electron and
$\langle \cdots \rangle$ is the expectation value with the
WP $\phi(\bfr,t)$.
They are derived from the time-dependent variational 
principle.
For many-electron wave functions
with non-orthogonal WPs, 
the equations become much
more complicated, which we shall leave out of the scope of this paper.

\section{Numerical Calculation and Discussion}

\subsection{Bond Formation and Excitation}

\begin{figure}[t]
    \centering
    \includegraphics[width=0.45\textwidth]{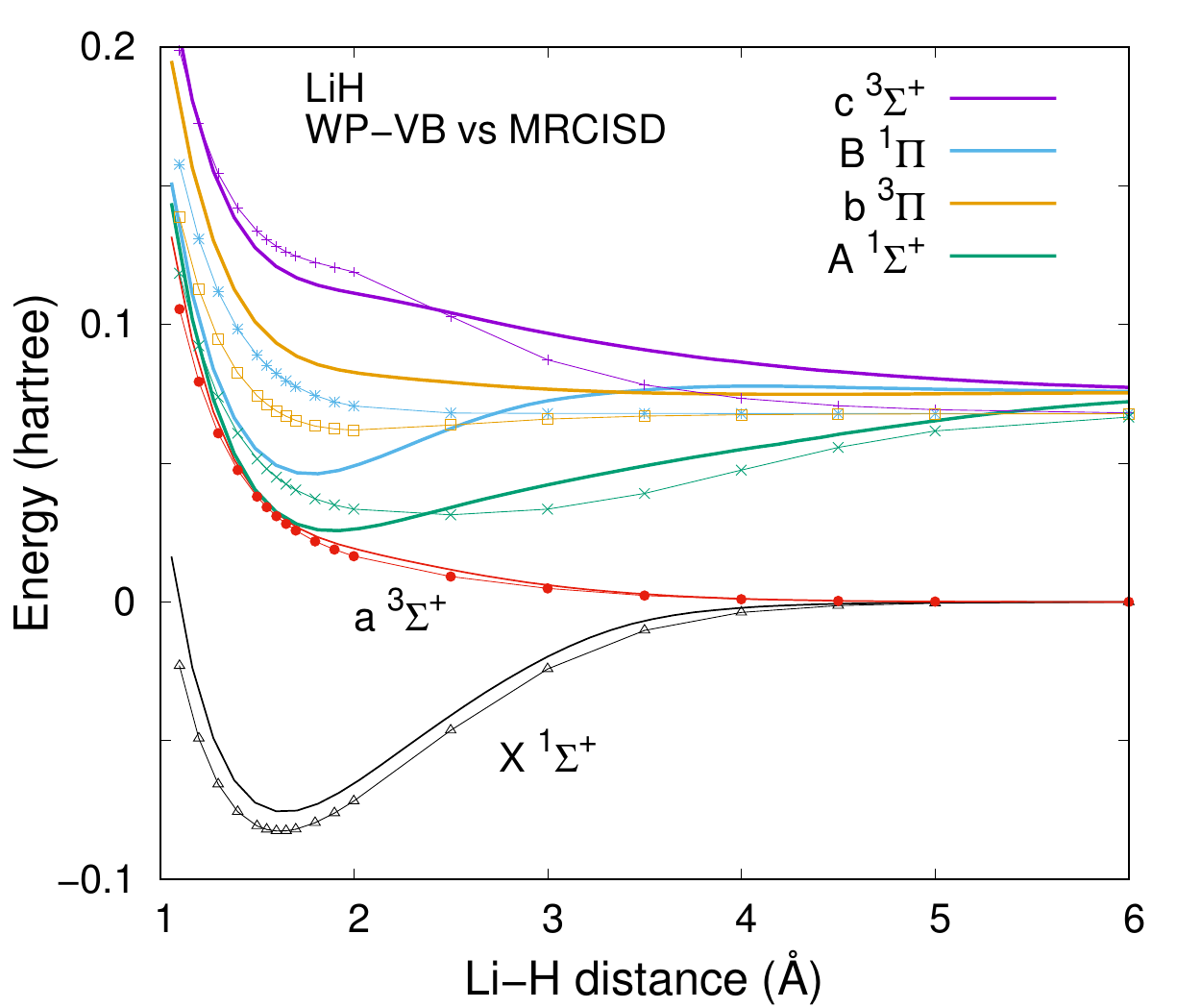}
    \caption{
        Potential energy curves of the ground and excited singlet and triplet
        states of LiH, 
        $X^{1}\Sigma^{+}, A^{1}\Sigma^{+}, B^{1}\Pi, 
        a^{3}\Sigma^{+}, b^{3}\Pi$, and $c^{3}\Sigma^{+}$.
        The solid lines are from the WP-VB calculation and the symbols
        are from the MRCISD calculation
        with the cc-pVDZ basis set.
    }
\end{figure}

Figure 1 displays
the computed potential energy curves of the ground and
excited states of LiH in spin singlet and triplet. 
The excited states were calculated by quantizing
the motion of the center of a WP 
that corresponds to the Li 2s electron:
we first constructed potential energy surfaces along displacements
of the WP centers (Fig. 2 to be discussed later)
and then numerically solved the time-independent Schr\"odinger equation
in a single-particle mean-field approximation.
For comparison, results from 
the multi-reference CI 
with single and double excitations (MRCISD) \cite{Schmidt1993} 
are included in the figure.
The agreement is quantitative
for the singlet and triplet ground states,
$X^{1}\Sigma^{+}$ and $a^{3}\Sigma^{+}$.
For the excited states, 
the correspondence appears semi-quantitative.
More elaborate calculations
that will be noted in the concluding section will improve the agreement.
Nevertheless, we proceed with the current framework as it
offers a simple intuitive picture for the electron excitation dynamics.

\begin{figure}[t]
    \centering
    \includegraphics[width=0.45\textwidth]{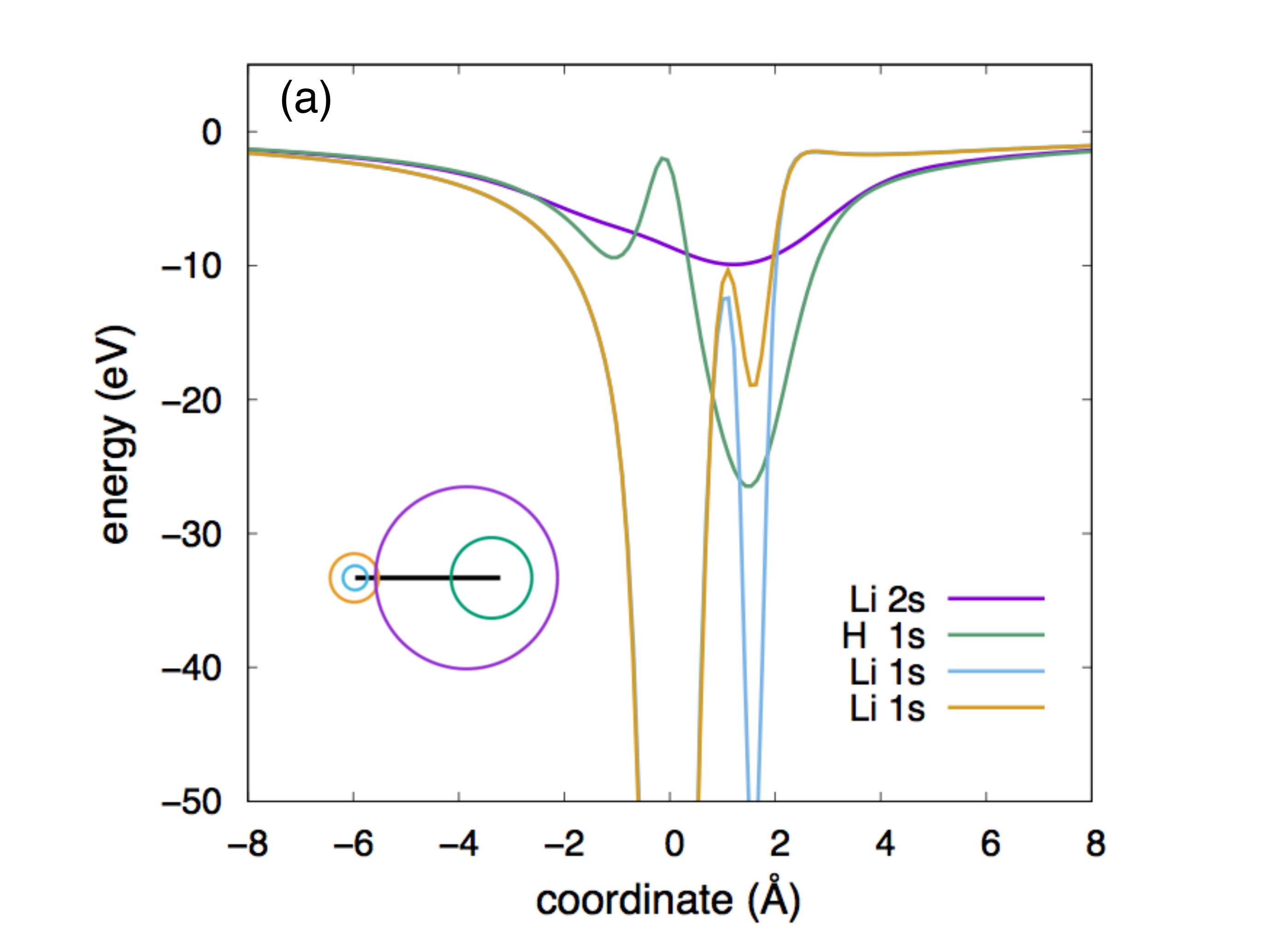}
    \includegraphics[width=0.45\textwidth]{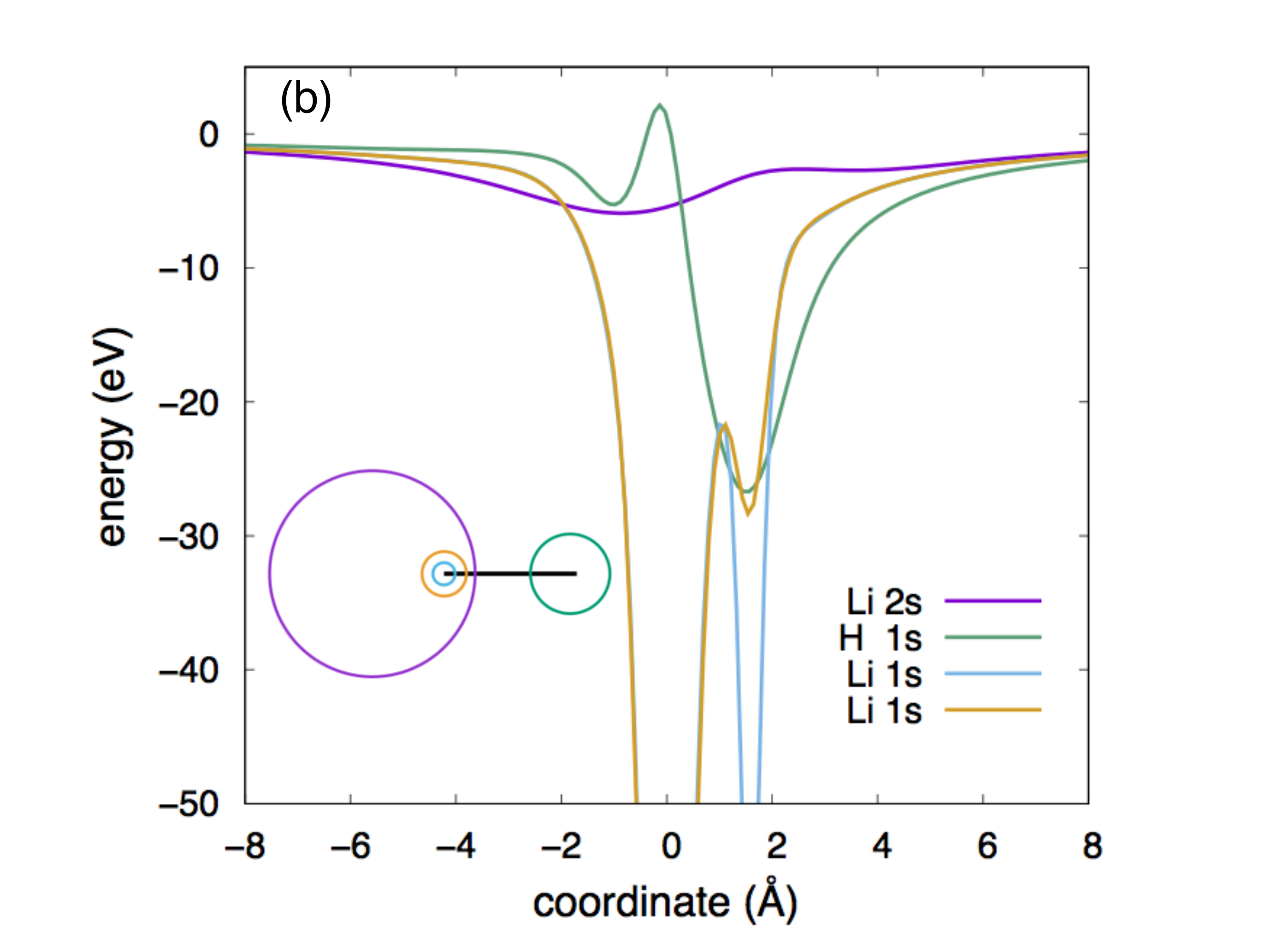}
    \caption{
        Potential energy curves for the motion of wave packet centers in
        (a) the singlet $X^{1}\Sigma^{+}$ state and 
        (b) the triplet $a^{3}\Sigma^{+}$ state.
        The insets are the wave packets represented by circles
        with the radius of the WP width $\rho$.
    }
\end{figure}

Figure 2 displays the potential energy curves for the WP motion along
the bond direction
in the ground singlet (a) and triplet (b) states. 
At the origin is the Li nucleus, and the proton is at 1.6 \AA.
In this calculation, the center of each WP was displaced 
while the other WPs are fixed, which means that
the potential energy of each WP was evaluated under the mean-field of others.
Figures 2(a) and (b) both indicate that the
two WPs that correspond to the Li 1s electrons are deeply bound and contracted
at the Li nucleus. The WP that corresponds to the H 1s electron is
bound to the proton with the potential well
shallower than those for the Li 1s electrons. The WP that corresponds
to the Li 2s electron is near the H nucleus in the singlet state,
whereas in the triplet state it is behind the Li nucleus seen from the H atom.
The inset displays the WPs by the circles of the radius $\rho$.
It has been demonstrated in Fig. 2 of Ref. \cite{Ando16} 
that these WP configurations correspond well
with the standard molecular orbitals.
The potential
well for the Li 2s WP is the shallowest in both singlet and
triplet states.
The difference of the curvature from the others
supports the single-electron treatment for the calculation of
the low energy excitations in Fig. 1.
Hereafter, we loosely denote
\lq Li 2s WP\rq\ etc. rather than
\lq WP that corresponds to the Li 2s electron\rq,
although 
the correspondence is only intuitive as the WPs are
not clumped at the nuclear center and the size is variable.

The potential energy curves along the displacement perpendicular to
the bond that determine the $\Pi$ excited states are simple symmetric single wells
(and thus were omitted from the figure) of varying depth. Here again,
the potential well for the Li 2s electron is the shallowest with the
smallest curvature that determines the lowest $\Pi$ states in
Fig. 1.

We note that
the WP widths  were fixed at the values optimized for the
ground state 
when constructing the potential energy curves in Fig. 2.
This is also the case in the calculation of excitation energies 
at each LiH distance in Fig. 1.
Alternatively, we might consider 
variation of the WP width for quantization.
We might also optimize adiabatically
the WP width along the variation of the WP center. Nonetheless,
the physical meanings of these procedures have not been clarified. 
A related study on the WP width variation in the initial-value represented propagator
has been reported \cite{Ando14}.

\subsection{Electron Dynamics and High-Harmonic Generation}

As described in Sec. 2, 
the computation of
single-particle dynamics is straightforward.
Here we focus on the Li 2s WP that is the most labile to the
external field,
and compute its response dynamics to a laser pulse of
time-dependent electric field,
\begin{equation}
E(t)=E_{0}\sin(\omega_{0}t)\sin^{2}(\pi t/\tau)
\hspace*{1em}
(0 \le t \le \tau).
\end{equation}
The parameters were taken from Ref. \cite{Sato15} that employed the time-dependent
complete-active-space CI (TD-CASCI); the frequency
$\omega_{0}$ corresponds to the wavelength of 750 nm and the duration
$\tau$ of three optical cycles, $\tau=3(2\pi/\omega_{0})\simeq7.51$
fs. We used weaker field intensity $E_{0}$ than in Ref. \cite{Sato15}; $-1.54$ and
$-1.48$ V/cm, that is, the laser intensity of $1.12\times10^{14}$ and
$1.08\times10^{14}$ W/cm$^{2}$, as these values were found to be 
just above and
below the threshold of ionizing trajectories
for the present model. 
The positive direction of the field was defined
to be that from Li to H.
Here, the Li-H bond length was set
to be 2.3 bohr following Ref. \cite{Sato15}.
At $t=0$, the LiH molecule is in the singlet ground state.
We added the Lorentz force from the electric
field to the equations of motion (\ref{eq:EOMqp}) for the WP center.

\begin{figure}[t]
    \centering
    \includegraphics[width=0.45\textwidth]{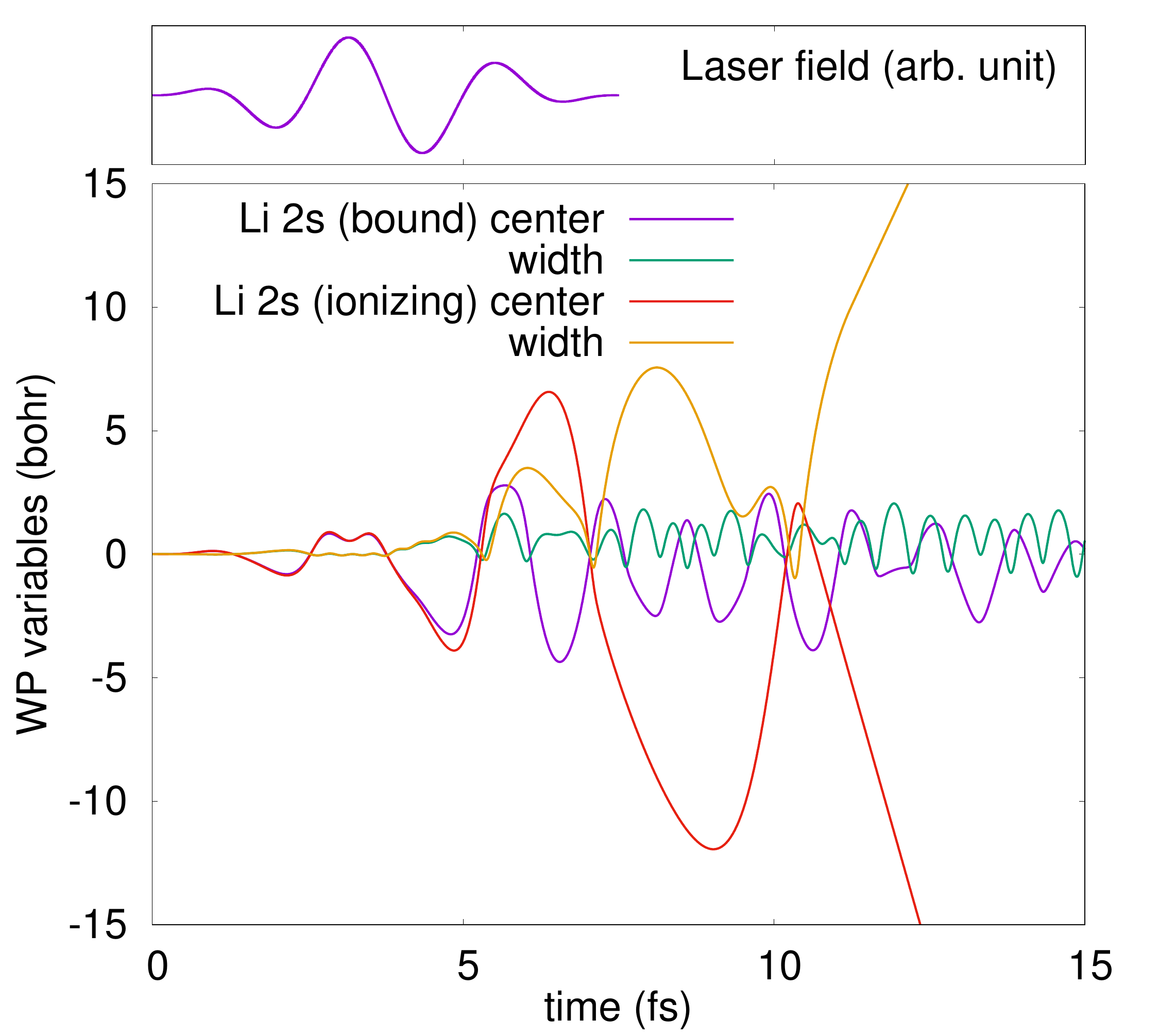}
    \caption{
        Trajectories of the wave packet center and width induced by
        the pulse laser field displayed in the upper panel.
    }
\end{figure}

Figure 3 displays the ionizing and non-ionizing (bound) trajectories 
of the WP center and width.
In the initial time up to $\sim$5.7 fs, both
the ionizing and bound trajectories of the WP center 
mostly follow the time profile of the laser
field, except the small hump at $t = 3.2$ fs. 
Then, the ionizing trajectory starts to oscillate in an
amplitude as large as 12 bohr, 
recombines to the molecule, and dissociates 
after 10 fs. Associated with this large amplitude oscillation
of the WP center,
the WP width broadens when the WP center departs from the molecule, and
spreads to the free electron state. In the bound trajectory,
the WP center remains around the molecule, but the oscillation amplitude
of $\sim$5 bohr is about twice the Li-H bond length of 2.3 bohr.

\begin{figure}[t]
    \centering
    \includegraphics[width=0.45\textwidth]{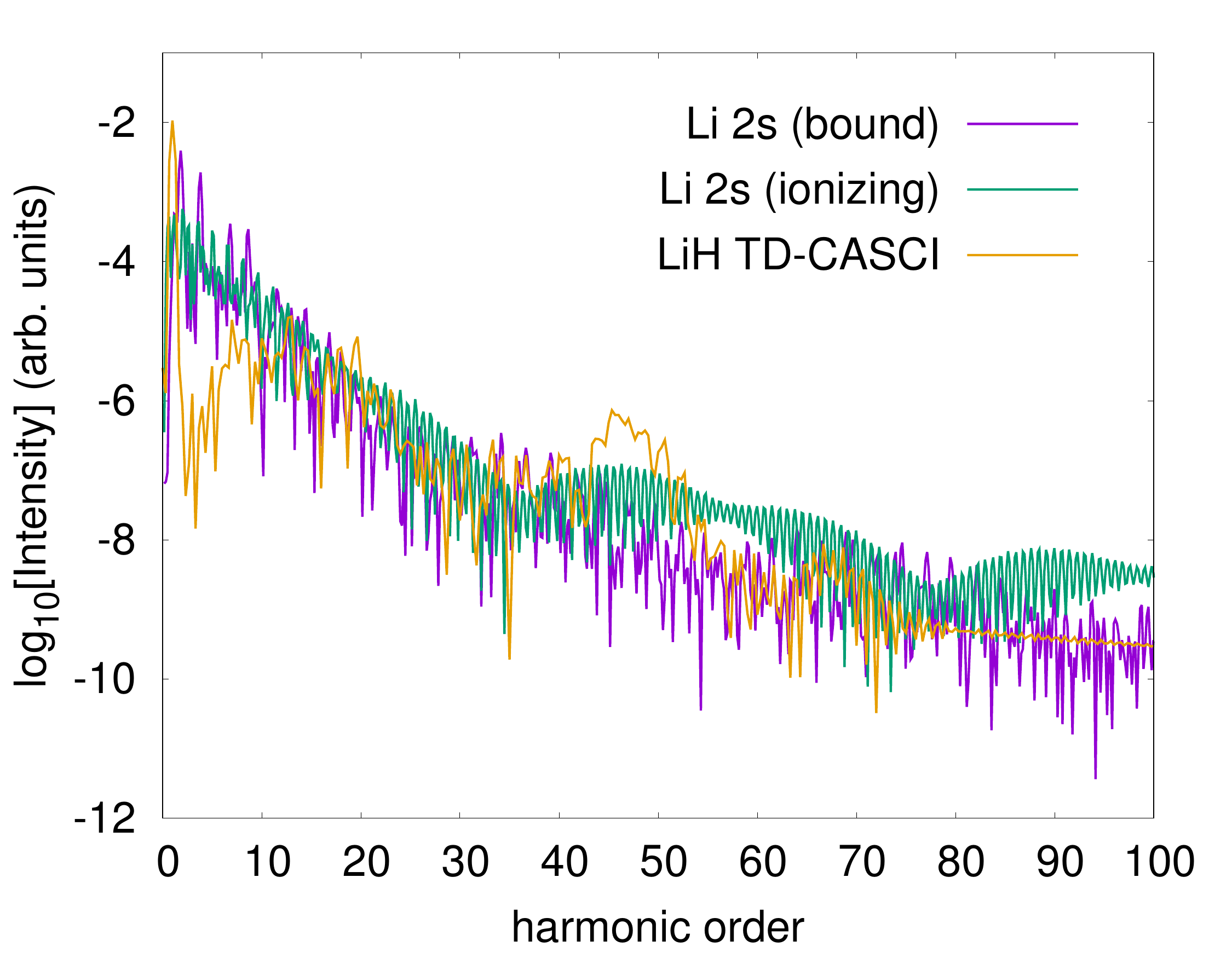}
    \caption{
        Fourier transform of the dipole acceleration
        that gives the HHG spectra.
        The abscissa is the harmonic-order $\omega/\omega_0$.
        The TD-CASCI data is from Ref. \cite{Sato15}.
    }
\end{figure}

Figure 4 displays the Fourier transformed 
dipole acceleration
that gives the high-harmonic generation (HHG) spectrum \cite{Baggesen11,Bandrauk13}.
For comparison, the result from 
the TD-CASCI of Ref. \cite{Sato15} is included.
The present results do not exhibit clear plateau and cut-off
that are considered to be the characteristics of the laser-induced HHG spectra.
These features have been discussed in terms of the ponderomotive energy of
field-induced dynamics and the three-step model of ion recombination
\cite{Krause92,Schafer93,Corkum93}.
Their absence in the present semiquantal calculations
could be due to the insufficient
account of the quantum interference in the ion recombination dynamics. 
However, other qualitative
features such as the alternating peak structure intense at the odd
harmonic orders and the spectral intensity extending up to a hundred
of harmonic order
were reproduced.

\section{Conclusion}

A model of floating and breathing electron wave packets
with the non-orthogonal valence-bond spin-coupling has been presented.
Although it is still in need of further developments,
its potential would have been clarified. 
We anticipate that the remaining tasks will
be more technical than fundamental. 
For instance, more flexible forms
of the localized WP than the spherical Gaussian should be examined.
The first candidate would be the ellipsoidal Gaussian function:
we have reported its use for nuclear WPs in the simulation
of liquid water \cite{Ono12}, but its application to many-electrons
involves
a numerical bottleneck in the evaluation of two-electron integrals.
We also need to establish the propagator theory with
the coherent-state representation and its implementation
in order to improve the accuracy of quantum dynamics.
Works on these are currently under way and will be reported
in due course.

\section{Acknowledgment}

The author is grateful to Professors Takeshi Sato and Kenichi Ishikawa
for providing their data in Fig. 4.
This work was supported by KAKENHI No. 26248009 and 26620007.

\end{document}